\documentclass[a4paper]{article}
\usepackage[dvipsnames]{xcolor}
\usepackage{multirow}
\usepackage{url}
\usepackage{hyperref}
\usepackage{lipsum}
\usepackage{INTERSPEECH2019}

\title{FeatherWave: An efficient high-fidelity neural vocoder with multi-band linear prediction}
\name{Qiao Tian, Zewang Zhang, Heng Lu, Ling-Hui Chen, Shan Liu}

\address{
  Tencent, China}
\email{\{briantian, zewangzhang, bearlu, nedchen, shiningliu\}@tencent.com}
\begin{document}

%
\maketitle
\newcommand\blfootnote[1]{%
\begingroup
\renewcommand\thefootnote{}\footnote{#1}%
\addtocounter{footnote}{-1}%
\endgroup}

\bmdefine{\bO}{O}
\bmdefine{\bC}{C}
\bmdefine{\bc}{c}
\bmdefine{\bo}{o}
\bmdefine{\bW}{W}
\bmdefine{\bmu}{\mu}
\bmdefine{\bQ}{Q}
\bmdefine{\bq}{q}
\bmdefine{\bw}{w}
\bmdefine{\bU}{U}
\bmdefine{\bu}{u}
\bmdefine{\bZero}{0}
\bmdefine{\bI}{I}
\bmdefine{\bR}{R}
\bmdefine{\bP}{P}
\bmdefine{\br}{r}
\bmdefine{\bmm}{m}
\bmdefine{\bsigma}{\sigma}
\bmdefine{\bSigma}{\Sigma}
\bmdefine{\bS}{S}
\bmdefine{\bA}{A}
\bmdefine{\ba}{a}
\bmdefine{\bD}{D}
\bmdefine{\bE}{E}
\bmdefine{\bF}{F}
\bmdefine{\bbf}{f}
\bmdefine{\bM}{M}
\bmdefine{\bg}{g}
\bmdefine{\bs}{s}
\bmdefine{\bpsi}{\psi}
\bmdefine{\bPsi}{\Psi}
\bmdefine{\bphi}{\phi}
\bmdefine{\bPhi}{\Phi}
\bmdefine{\bPi}{\Pi}
\bmdefine{\bpi}{\pi}
\bmdefine{\bLambda}{\Lambda}
\bmdefine{\blambda}{\lambda}
\bmdefine{\bB}{B}
\bmdefine{\bb}{b}
\bmdefine{\bl}{l}
\bmdefine{\bd}{d}
\bmdefine{\bD}{D}
\bmdefine{\bG}{G}
\bmdefine{\bp}{p}
\bmdefine{\bxi}{\xi}
\bmdefine{\bmeta}{\eta}
\bmdefine{\bzeta}{\zeta}
\bmdefine{\bk}{k}
\bmdefine{\bX}{X}
\bmdefine{\bx}{x}
\bmdefine{\bY}{Y}
\bmdefine{\by}{y}
\bmdefine{\bH}{H}
\bmdefine{\bh}{h}
\bmdefine{\bV}{V}
\bmdefine{\bv}{v}
\bmdefine{\bn}{n}
\bmdefine{\bz}{z}
\bmdefine{\bZ}{Z}
\bmdefine{\bOmega}{\Omega}
\bmdefine{\bomega}{\omega}
\bmdefine{\bGamma}{\Gamma}
\bmdefine{\bgamma}{\gamma}
\def\diag{\mathrm{diag}}
\def\idiag{\mathrm{diag}^{-1}}
\def\tr{\mathrm{tr}}
\def\exp{\mathrm{exp}}
\def\sign{\mathrm{sign}}

\begin{abstract}
\label{sec:abs}


In this paper, we propose the FeatherWave, yet another variant of WaveRNN vocoder combining the multi-band signal processing and the linear predictive coding.
The LPCNet, a recently proposed neural vocoder which utilized the linear predictive characteristic of speech signal in the WaveRNN architecture, can generate high quality speech with a speed faster than real-time on a single CPU core. 
However, LPCNet is still not efficient enough for online speech generation tasks. 
To address this issue, we adopt the multi-band linear predictive coding for WaveRNN vocoder. 
The multi-band method enables the model to generate several speech samples in parallel at one step. 
Therefore, it can significantly improve the efficiency of speech synthesis.
The proposed model with $4$ sub-bands needs less than $1.6$ GFLOPS for speech generation.
In our experiments, it can generate $24$ kHz high-fidelity audio $9$x faster than real-time on a single CPU, which is much faster than the LPCNet vocoder.
Furthermore, our subjective listening test shows that the FeatherWave can generate speech with better quality than LPCNet.
\end{abstract}
\noindent\textbf{Index Terms}:  WaveRNN, LPCNet, multi-band, linear prediction
%
%
\section{Introduction}
\label{sec:intro}

In recent years, the quality of text-to-speech (TTS) has been significantly improved by neural vocoders such as WaveNet~\cite{van2016wavenet}, Parallel WaveNet ~\cite{oord2017parallel}, WaveRNN ~\cite{kalchbrenner2018efficient}, LPCNet ~\cite{valin2019lpcnet}, etc.
These neural vocoders are usually used in sequence-to-sequence acoustic models, e.g. Tacotron 2 ~\cite{shen2018natural} and DurIAN ~\cite{yu2019durian}, to achieve generating human-like speech.
The WaveNet vocoder, which is the state of the art model, can generate high-fidelity audio but is hard to deploy for real time services because of its huge computational complexity. 
The flow based neural vocoders, such as Parallel WaveNet~\cite{oord2017parallel}, Clarinet~\cite{ping2018clarinet}, WaveGlow~\cite{prenger2019waveglow}, are more practicable since they can perform parallel generation on GPU devices.
However, these models often suffer from phase issues since the causality prior is ignored.
Therefore the generated speech usually sounds muffled compared with the original auto-regressive WaveNet.
Generative Adversarial Network (GAN)~\cite{goodfellow2014generative} has been adopted to address these issues in Parallel WaveNet ~\cite{tiangenerative,yamamoto2019probability}.

Recently, efficient RNN based sequential neural vocoders, such as WaveRNN, LPCNet and Multi-band WaveRNN~\cite{yu2019durian}, have been proposed for improving the performance of neural TTS system.
The proposed LPCNet is the most lightweight neural vocoder currently, which integrates WaveRNN structured neural synthesis techniques with linear prediction.
Meanwhile, an improved sampling strategy, as well as the pre-emphasis prior to $\mu$-law  quantization is introduced for achieving good quality under a small model size.
Different from separately predicting the coarse and fine parts of the discretized speech signal in WaveRNN, LPCNet replaces the dual softmax output layer with a single softmax output layer on the $8$-bit $\mu$-law quantized signal with pre-emphasis.
As a result, the LPCNet can produce $16$ kHz high quality speech with a complexity less than $3$ GFLOPS, which significantly improved the speed of speech synthesis system.
On the other hand, the Multi-band WaveRNN is a variant of WaveRNN, which integrates multi-band strategy into WaveRNN based neural vocoder. Compared with WaveRNN, Multi-band WaveRNN can produce multi samples at one sequential step in parallel.

However, neural TTS systems with low computational complexity are very important for practical applications. 
As reported in ~\cite{valin2019lpcnet} and ~\cite{yu2019durian}, both LPCNet and Multi-band WaveRNN can not be $5$x faster than real-time when producing $24$ kHz high quality speech with a single CPU core, which means that the latency of synthesizing one second speech could be more than $200$ms.
Furthermore, there are many applications that require synthesizing speech on edge-devices, such as mobile phones with very limited computational capacity. 
For this purpose, we propose the FeatherWave vocoder, which merges multi-band processs to LPCNet framework. 
This makes it possible to match the quality of the state of the art neural vocoder WaveNet with significantly smaller computational load.

The contributions of this paper are summarized as follows:
(1) We propose the multi-band (MB) linear prediction (LP) based FeatherWave vocoder.
Firstly, we adopt the multi-band signal processing into the LPCNet framework. 
Then, we combine the $\mu$-law quantization with MB-LP for efficiently modeling the discretized speech signal.
Benefiting from the MB-LP process, the complexity of the proposed model is significantly reduced compared with the conventional LPCNet.
(2) We demonstrate that the proposed FeatherWave can be $10$x faster than real-time on two CPU cores by using our engineered streaming inference kernel when generating $24$ kHz high-fidelity speech, which achieved a mean opinion score (MOS) of 4.55 in our subjective listening test.

We organize the rest of the paper as follows:
in Section~\ref{sec:re}, we will briefly review the lightweight RNN based neural vocoder, such as Multi-band WaveRNN and LPCNet.
Then the proposed method will be given in Section~\ref{sec:FeatherWave}.
The evaluation of results will be presented in Section~\ref{sec:exps}.
Lastly in Section~\ref{sec:cons}, conclusions and future work are presented.
\section{Related work}
\label{sec:re}

\subsection{Multi-band WaveRNN}
\label{sec:mwrnn}

Compared with Subscale WaveRNN ~\cite{kalchbrenner2018efficient}, which can generate multi samples per step with a subscale dependency scheme,
Multi-band WaveRNN exploits multi-band generation strategy with the technique of subband~\cite{okamoto2018improving,okamoto2018investigation} to improve generation speed. 
It predicts all subband signal simultaneously through a multiple softmax output layer in a single recurrent step in WaveRNN.
By using this variant of WaveRNN, the length of generated sequence can be down-sampled by a factor of $N_b$ (the number of frequency bands). 
As a result, the total computational cost can be reduced to approximately $3.6$ GFLOPS ~\cite{yu2019durian}.
Before model training, the original waveform signal ${x=\{x_1, \dots, x_T\}}$ should be down-sampled by $N_b$ invertiable analysis filters into ${N_b}$ subbands waveforms $g = \{g^{b}\}, b = 1, \dots, N_b$, where ${g^{b}=\{g_1^{b}, \dots, g_{T/N_b}^{b}\}}$.
The joint probability of multi-band signal can be factorised as a product of conditional probabilities of subband signals as described as
\begin{equation}
\label{mwrnn}
  p(g) = \prod _{n=1}^{T/N_b} p( g_n | g_1, g_2 , \dots, g_{n-1}),
\end{equation}
where the conditional probability can be modeled by a recurrent neural network (RNN).

\subsection{LPCNet}
\label{sec:lpcnet}

The LPCNet makes effort to reduce the computational load of each generation step benefiting from the classical technique of linear prediction.
Similar to GlotNet \cite{juvela2019glotnet} and ExcitNet \cite{song2019excitnet} which use the WaveNet to capture the glottal excitation signal, 
the LPCNet models the discretized excitation signal of LPC filters with a WaveRNN for efficient generation.
Instead of open-loop filtering approaches~\cite{juvela2018speaker}, 
LPCNet preforms as a closed-loop synthesis of predicting sample $x_t$ by conditioning on the previously sampled excitation $e_{t-1}$ and current prediction $p_t$, 
which can improve the quality of generated speech. 

\section{The proposed method}
\label{sec:FeatherWave}

In this section, we present the proposed variant of WaveRNN vocoder, FeatherWave, which further improves the speed of audio generation with multi-band process and maintains the advantages of the LP-structure as LPCNet.
Firstly, we introduce the MB-LP framework, which extends the process of the LP coding to multi-band signal.
Then, we propose the FeatherWave vocoder which applies the MB-LP framework into the conventional neural vocoder.

\subsection{Multi-band Linear Prediction}
\label{sec:mblp}

For the purpose of utilizing linear prediction to obtain good quality and multi-band to speed up synthesis, we introduce multi-band linear prediction (MB-LP) in the proposed model. 
By adopting LP analysis on multi-band waveform signal, $M$ order linear prediction coefficients of each sub frequency band, ${\alpha{_k^b}}$, 
can be extracted from the corresponding frequency bins of mel-spectrogram frame.
The $b$-th subband signal $g^b$ is down-sampled from the original signal $x$ by invertible analysis filters.
Under the LP assumption, the corresponding predicted signal $p_n^b$ and excitation (prediction residual) $e_n^b$ of $b$-th band can be computed as follows:

\begin{equation}
\label{mblp_1}
  p_n^b = \sum _{k=1}^{M} \alpha _k^b ~ g_{n-k}^b,
\end{equation}
\begin{equation}
  \label{mblp_2}
  g_n^b = p_n^b + e_n^b.
\end{equation}

\begin{figure}[ht]
\centering
	\begin{minipage}[b]{1.0\linewidth}
		\centerline
		\centerline{\includegraphics[width=8cm, height=8cm]{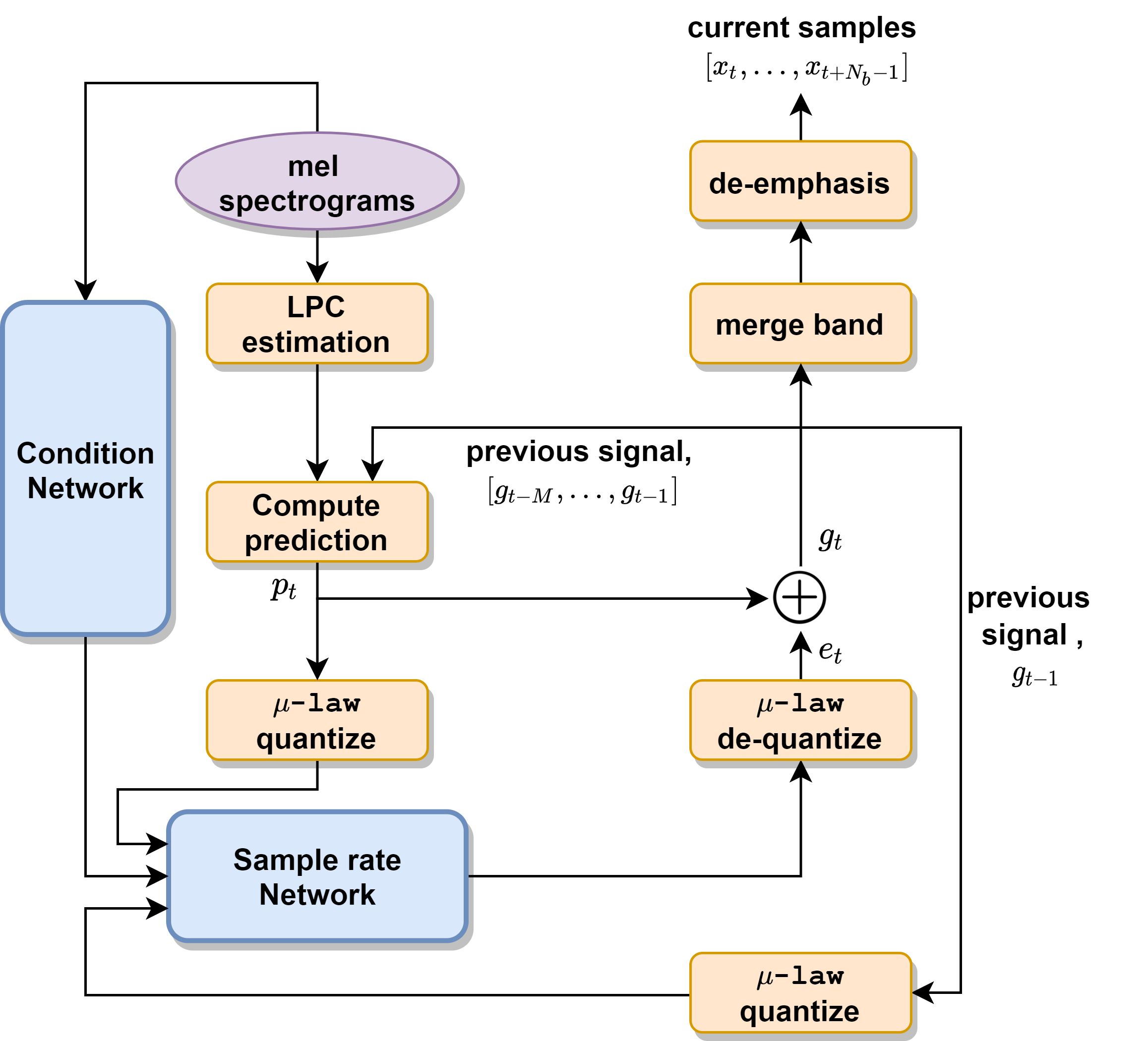}}
	\end{minipage}
	\caption{Block diagram of the proposed FeatherWave vocoder.}
	\label{fig:fw_arc}
\end{figure}

\subsection{FeatherWave}
\label{sec:fw}
In our proposed FeatherWave vocoder, MB-LP is introduced into the conventional WaveRNN vocoder as illustrated in Fig.~\ref{fig:fw_arc}. 
It consists of a condition network that operates on input frames of mel spectrograms 
and a sample rate network which produces ${N_b}$ samples with a multi dual softmax output layer. 
Similar to the original WaveRNN, the sampling network firstly predicts coarse part of excitation signal and then computes fine part by conditioning on the predicted coarse signal. 
As indicated in Eq.~\ref{mblp_2}, the subband signal is predicted from the network output excitation and linear predicted signal, which is linearly predicted from previous output signal as show in Eq.~\ref{mblp_1}.
As illustrated in Fig.~\ref{fig:fw_arc}, the merge band operation is applied, by using synthesis filters, to reconstruct original waveform signal from the predicted signal of subbands.
In this paper, only mel spectrograms, which are widely used in neural TTS systems, are adopted as input conditional features.

\subsubsection{Discretized Multi-band Linear Prediction}
\label{ssec:dmblp}

In LPCNet, a first-order pre-emphasis filter $E(z)=1-\alpha z^{-1}$ is applied to training data. 
This pre-emphasis makes it possible to model $8$-bit $\mu$-law discretized signal with high quality.

As an obvious extension of using this technique to help model learn and generate more efficiently, we also apply this pre-emphasis filter to training signal firstly, and then $\mu$-law quantize all subbands signals after MB-LP process. 
Similar to LPCNet, we can model $\mu$-law discretized signal using smaller model and achieve high-fidelity synthesis with the proposed MB-LP framework. 
For trading off quality against model size, we adopt $10$-bit $\mu$-law quantization for each subband signal in the FeatherWave.

\subsubsection{Condition Network}
\label{ssec:cn}

For neural vocoder, the intelligibility of generated speech is much sensitive to the structure of condition network.
In FeatherWave, instead of using bi-directional RNN, we adopted a stack of convolutional layers as the condition network for the purpose of streaming inference.
Specifically, the local acoustic features are firstly operated by five $1 \times 3$ convolution layers so the sample rate network can obtain enough receptive field.
We adopt exponential linear unit (ELU) activation after every convolutional layer for more stable training. 
In order to match the sampling rate of target signal, the outputs of condition network are simply repeated by $f$ times before passed into sample rate network.
As $h$ denotes hop size, the number of repetitions is $f = h / N_b$.

\subsubsection{Sample Rate Network}
\label{ssec:sn}

In the sample rate network, predictions computed from linear prediction are conditioned for the manner of closed-loop synthesis by following the method in LPCNet. 
As a result, the predictions perform as reference signal to compute excitations.
This can enhance the performance of model.
Besides, the up-sampled features from the output of condition network and the previous generated signal are used as well. 
All discretized signals are passed into a trainable embedding layer before fed into a GRU cell.
Similar to the WaveRNN vocoder, we use dual softmax layer to predict coarse and fine parts of the discretized signal sequentially after a GRU and affine layers. 
A block sparse pruning~\cite{narang2017block} strategy is adopted to sparsify the parameters in the GRU layer for the purpose of speeding up inference.
The output of the affine layer is passed into multiple softmax output layers to predict all subband excitations simultaneously. 
The parameters of model are optimized to minimize the negative log-likelihood (NLL) loss at the training phase.

\subsubsection{Generation Method}
\label{ssec:ge}

In typical lightweight neural vocoder where small model is adopted, 
it is necessary to adjust the sharpness of the output distributions to avoid noise caused by the random sampling process and achieve better quality.
In FFTNet~\cite{jin2018fftnet} and iLPCNet~\cite{hwang2020improving}, lowering temperature in the voiced region with a constant factor is exploited for such purpose.
Rather than using voiced information, LPCNet adopts pitch correlation to adjust the temperature factor. 
Furthermore, the distribution is subtracted with a constant threshold $T$ to prevent impulse noise caused by low probabilities.

Since only mel-spectrograms are used in condition network, we explore the technique of distribution subtraction carefully for better performance.
We observed that a temperature $T=0.02$ produced good results in the trade-off quality against artifact in generated speech. The subtraction is only performed on the distribution of fine part, which is given as follow:
\begin{equation}
\label{p_f}
  P^{'}_f(e_t) = \mathcal R(\max[P_f(e_t) - T, 0]),
\end{equation}
where $P_f(e_t)$ denotes the distribution of fine part, 
and $\mathcal R(\cdot)$ denotes the normalizing operator.
  
\subsection{Two-stage Sparse Pruning}
\label{sec:pst}
In ~\cite{kalchbrenner2018efficient}, a GRU with block sparse weights is vital for achieving fast inference in neural vocoders. 
In this work, in order to improve the performance of block sparse strategy, we apply a novel two-stage sparse pruning (TSSP) method to achieve high sparsity ratio in GRU weights.
 
In the conventional block sparsity pruning methods, a high sparsity ratio (above 40\%) usually degrades the model  performance
as mentioned in ~\cite{yao2019balanced}.
In practice, high sparsity ratio usually hurts the speech quality of neural vocoders, although it could speed up the inference.
To address this problem, we adopt a two-stage sparse pruning strategy, which consists of warming-up stage and increasing stage.
Firstly, we train sparse model with a warming-up sparsity ratio which is $50$\% in our configuration to avoid hurting performance of model in warming-up stage. 
In the increasing stage, we increase the sparsity ratio progressively by loops to reach the target sparsity ratio, e.g.  increasing $10$\% sparsity ratio in a loop.
We maintain the sparsity ratio with a constant iterations 
after the warming-up sparsity ratio or the target sparsity ratio of every loop in increasing stage is reached.


\section{Experiments}
\label{sec:exps}

\subsection{Data Set}
\label{ssec:data}
In our experiments, we used a Mandarin corpus of 20 hours of recordings, which were recorded by a professional broadcaster.
The data we split into a training set and a test set. 
About $18$ hours of recordings were used for model training and the rest were used for testing.
All the recordings were down-sampled to $24$ kHz sampling rate with $16$-bit format.
The 80 order mel-spectrograms were extracted as the conditions for all neural vocoders in our experiments with the method mentioned in ~\cite{shen2018natural}.
 
\subsection{Experimental Setup}
\label{ssec:vocoders}
To demonstrate that the proposed model accelerates speech synthesis without degrading the speech quality, we chose LPCNet, which is open-sourced and is known as the fastest high-quality neural vocoder, as the baseline.
In the LPCNet baseline system, we used the open-sourced implementation \footnote{\url{https://github.com/mozilla/LPCNet/}}
based on the commit $3a7ef33$ and the configuration was exactly the same as its original version.
A $384$-dimensional GRU layer with $90$\% sparsity ratio before the $16$-dimensional dense GRU layer was used.
Since the LPCNet open-sourced implementation can only generates $16$ kHz audio, we down-sampled the generated speech of the proposed model for a fair comparison. 
The original $24$ kHz speech of our model is included in the comparison as well.
In order to observe the gap between the proposed model and the state-of-the-art neural vocoder, a WaveNet with mixture of logistic (MoL) output layer was also adopted for comparison.
For robustness and stability, we chose the MoL WaveNet variant~\cite{tian2019tencent} and all the configurations were the same as mentioned in ~\cite{tian2019tencent}.

In the proposed FeatherWave vocoder,
conv1d layers with kernel size 1$\times$3 and channel size 256 were used in the condition network.
In sample rate network, the final sparsity ratio is set as $90$\% in the GRU with 384 hidden units.
The dimension of affine layer is $128$. 
The embedding size for discretized signal is $16$.
In this work, we used $4$ bands and $10$-bit $\mu$-law quantization for dual softmax layers, therefore the output dimension of the last FC layer before softmax layer was $128$.
For modeling and reconstructing on subband signal, we followed the design of analysis filters and synthesis filters in~\cite{nguyen1994near}.
Instead of adopting cepstrums~\cite{valin2019lpcnet}, the LP coefficients were estimated from the mel-spectrograms as in ~\cite{korostikstc}.

In the training phase, the Adam~\cite{kingma2014adam} optimizer was adopted with a learning rate of $0.001$.
The proposed model was trained on a single GPU with mini-batch size of 1536 samples.
The weights of the neural vocoders were randomly initialized with fixed random seed and all the networks were trained with $1200$k iterations. 
In the two-stage sparse pruning of FeatherWave, the target sparsity ratio of the warming-up stage was $50$\% with $300$k sparse iterations and continued to the increasing stage after maintaining the current sparsity with $100$k iterations.
In every loop of increasing stage, the sparsity ratio was increased by $10$\% with $100$k iterations and maintaining the current sparsity with another $100$k iterations.
After four loops in increasing stage, the total iterations reached $1200$k and the final sparsity ratio was $90$\%, which is same as in the LPCNet. 
The blocks with size $16\times1$ were adopted in our pruning experiments.

\subsection{Synthesis Speed}
\label{ssec:speed}

\begin{table}
\centering
\caption{The synthesis speed over real-time of the baseline model LPCNet and the proposed FeatherWave for two sampling rate ($16$ kHz and $24$ kHz) speech.}
\begin{tabular}{ccc}
\toprule
{\textbf{syn. speed}} & {\textbf{single core}} & {\textbf{two cores}} \\
\midrule
LPCNet & 5.7x & {-} \\
\textbf{FeatherWave (16k)} & \textbf{12.1x} & \textbf{15.5x} \\
\textbf{FeatherWave (24k)} & \textbf{9.2x} & \textbf{10.8x} \\
\bottomrule
\end{tabular}
\label{table:com}
\end{table}

We estimated the computational complexity of different vocoders firstly for revealing the speedup of our proposed FeatherWave vocoder.
The main complexity of FeatherWave comes from one sparse GRU and four fully-connected layers. 
We compute it following the method in ~\cite{valin2019lpcnet}, which is given by:
\begin{equation}
\label{Complexity}
  C = ({3d}{N^2_G} + {N_G}\cdot{N_F} + {2N_F}\cdot{Q}\cdot {N_B}) \cdot {2F_S} / {N_B},
\end{equation} 
where $N_G$ is the size of the sparse GRU, d is the density of the sparse GRU, $Q$ is the root of the number of $\mu$-law levels, $N_F$ is the width of affine layer connected with final fully-connected layer, $N_B$ is the number of frequency bands, and $F_S$ is the sampling rate. 
In our experiments, we set $N_G$ = $384$, $d$ = $0.1$, $Q$ = $32$, $N_F$ = $128$ and $N_B$ = $4$ for ${F_s}$ = $16000$.
Therefore, a total complexity of FeatherWave is approximately $1.6$ GFLOPS, which is much smaller than $2.8$ GFLOPS in the conventional LPCNet.

The synthesis speeds over real-time of different vocoders are listed in Table ~\ref{table:com}.
All the speed tests were performed on the Intel Xeon Platinum 8255C CPU. 
The results show that merging multi-band into LPCNet framework can bring about $2$x speedup when generating $16$ kHz speech.
When producing high-fidelity $24$ kHz speech, FeatherWave can be $10$x faster than real-time using our engineered multi-thread inference kernel on two CPU cores.
Additionally, our implementation of Parallel WaveNet \cite{tiangenerative} requires 8 cores to achieve the similar synthesis speed.

\subsection{Evaluations}
\label{ssec:mos}

Firstly, subjective evaluation was conducted to evaluate the MOS of perceptual quality of the proposed FeatherWave vocoder.
In order to perform fair comparison, we randomly selected 40 utterance from test set for MOS testing and 30 native Mandarin speakers participated in the listening test. 

The results\footnote{A subset of generated samples can be found at the following URL:\\\url{https://wavecoder.github.io/FeatherWave/}} of the subjective MOS evaluation is presented in Table ~\ref{table:mos}.
The results show that the proposed FeatherWave can generate high quality 16 kHz speech with a slightly better MOS than the LPCNet. 
And when producing high-fidelity speech at higher sampling rate (24 kHz), the proposed FeatherWave achieves a MOS with a small gap to the powerful MoL WaveNet, which consists of 24 dilated conv1d layers. 
Since we use mel-spectrograms to extract the LP filters, the proposed model doesn't depend on pitch extraction.
The model has less artifact in the generated speech and is easy to build a neural TTS system instead of LPCNet.
Furthermore, our model can produce less quantization noise and fidelity loss than LPCNet as $10$-bit $\mu$-law quantization with dual softmax layer is used instead of $8$-bit one. 

\begin{table}
\centering
\caption{Mean Opinion Score (MOS) with $95\%$ confidence intervals for different vocoders.}
\begin{tabular}{cc}
\toprule
{\textbf{Model}} & {\textbf{MOS on speech quality}} \\
\midrule
LPCNet & 4.48 $\pm$ 0.04 \\
FeatherWave (16k) & 4.51 $\pm$ 0.03 \\
\textbf{FeatherWave (24k)} & \textbf{4.55 $\pm$ 0.03} \\
MoL WaveNet & 4.58 $\pm$ 0.02 \\
\bottomrule
\end{tabular}
\label{table:mos}
\end{table}

We also investigated the effectiveness of two-stage sparse pruning method by objective NLL results.
Lower NLL usually indicates better quality of the neural vocoder generated speech~\cite{kalchbrenner2018efficient}. 
It is obviously observed from the results in Table ~\ref{table:pst} that the model got lower NLL compared with the baseline model after using the proposed two-stage sparse pruning method, which could lower the probability of bad choice in sparse pruning compared with the conventional pruning methods. Finally, we got better speech quality in FeatherWave with this improvement.

\begin{table}
\centering
\caption{FeatherWave NLL results on different sparse strategies.
All the experiments were conducted on the same sparsity ratio, 90\%.}
\begin{tabular}{cc}
\toprule
{\textbf{Method}} & {\textbf{NLL}} \\
\midrule
FeatherWave {\textbf{w/o TSSP}} & 4.14 \\
\textbf{FeatherWave w/~ TSSP} & \textbf{4.07} \\
\bottomrule
\end{tabular}
\label{table:pst}
\end{table}

\section{Conclusions and future work}
\label{sec:cons}

In this work, we proposed the FeatherWave vocoder which applies the MB-LP method to the conventional RNN based neural vocoder, such as WaveRNN. 
For faster generation and utilizing the linearity of the LP filters, we merged multi-band into LPCNet framework which only conditioned on mel spectrograms.
Furthermore, we also make other contributions, such as the discretized multi-band linear prediction and two-stage sparse pruning.
Our experimental results indicated that the proposed FeatherWave can further reduce the computational cost at speech generation and get higher speech quality compared with the conventional neural vocoders. 

In future work, we will investigate FeatherWave with low bit and balanced sparsity~\cite{yao2019balanced} pruning training method for deploying on edge-devices.

\section{Acknowledgments}
\label{sec:ack}
The authors would like to thank Yi Xie and Ciyong Chen in IAGS, Intel Asia-Pacific Research \& Development Co Ltd.. 
These two members in Intel not only provided the guidance on how to get good performance on the Intel(R) Xeon(R) Scalable Processors, 
but also helped to optimize/validate our algorithm with Intel(R) Deep Learning Boost using bfloat16 (BF16) format on the upcoming hardware.

\vfill\pagebreak

\bibliographystyle{IEEEbib}
\bibliography{refs}
\end{document}